\documentclass[pra,twocolumn,floatfix]{revtex4}

\usepackage{graphicx}
\usepackage{bm}
\usepackage{amsmath}

\newcommand{\be}{\begin{equation}}
\newcommand{\ee}{\end{equation}}
\newcommand{\ba}{\begin{eqnarray}}
\newcommand{\ea}{\end{eqnarray}}

\begin{document}

      \title{Algorithmic approach to adiabatic quantum optimization}

      \author{Neil G. Dickson}
      \author{Mohammad H.~Amin}
      \affiliation{D-Wave Systems Inc., 100-4401 Still Creek Drive,
      Burnaby, B.C., V5C 6G9, Canada}
\date{\today}

\begin{abstract}

It is believed that the presence of anticrossings with exponentially small
gaps between the lowest two energy levels of the system Hamiltonian,
can render adiabatic quantum optimization inefficient. Here, we present
a simple adiabatic quantum algorithm designed to eliminate
exponentially small gaps caused by anticrossings between eigenstates
that correspond with the local and global minima of the problem
Hamiltonian. In each iteration of the algorithm, information is
gathered about the local minima that are reached after passing the
anticrossing non-adiabatically.  This information is then used to
penalize pathways to the corresponding local minima, by adjusting the
initial Hamiltonian. This is repeated for multiple clusters of local
minima as needed. We generate 64-qubit random instances of the maximum
independent set problem, skewed to be extremely hard, with between
$10^5$ and $10^6$ highly-degenerate local minima.  Using quantum Monte
Carlo simulations, it is found that the algorithm can trivially solve
all the instances in $\sim 10$ iterations.

\end{abstract}
\maketitle

\section{Introduction}

Adiabatic quantum computation (AQC) \cite{Farhi01} is an important
paradigm for universal quantum computation \cite{Aharonov,Lidar}. In a
simple quantum adiabatic algorithm, the Hamiltonian of the system is
written as
 \be
 H= A(s) H_B + B(s) H_P, \qquad 0 \le s \le 1, \label{H}
 \ee
with $A(0)\gg B(0)$ and $A(1)\ll B(1)$. Here, $s=t/t_f$ is the
dimensionless time with $t_f$ being the total evolution time. The
system starts from a known ground state of $H_B$ at $t=0$ and ideally
ends in the ground state of $H_P$ at $t=t_f$, which is the solution to
a problem. To ensure that the system ends up in the final ground state
with high fidelity, the evolution should be very slow (adiabatic). In a
closed system, the total computation time is related to the minimum
energy gap $g_m$ between the ground state and first excited state of
the Hamiltonian. For problems with very small gap at $s=s^*$, a
two-state approximation near the anticrossing yields the success
probability $P_f = 1-e^{-t_f/t_a}$, with the adiabatic time scale
\cite{Amin08}
 \ba
 \left. t_a = \frac{4\hbar}{\pi}\frac{\langle
 0|dH/ds|1\rangle}{g_m^2} \right|_{s=s^*} \label{ta}
 \ea
where, $|0\rangle$ and $|1\rangle$ denote the ground and first excited
states. To determine the computational complexity of AQC, therefore,
one needs to know how $g_m$ scales with the problem size. This is
indeed not an easy task, although analytic calculations have been
possible for a few special examples
\cite{Aharonov,vanDam01,RolandandCerf,Znidaric06}.

An important subset of quantum adiabatic algorithms is adiabatic
quantum optimization (AQO), for which the final Hamiltonian $H_P$ is
diagonal in the computation basis. The desired final ground state of
the system is therefore the classical global minimum of $H_P$. In this
paper, we restrict ourselves to $H_P$ and $H_B$ of the form
 \ba
 H_P= \sum_i h_i \sigma^z_i + \sum_{i,j} J_{ij} \sigma^z_i\sigma^z_j,
 \quad \,
 H_B = - \sum_i \Delta_i \sigma^x_i, \label{HPB}
 \ea
where $\sigma^{x}_i$ and $\sigma^{z}_i$ are Pauli matrices, and $h_i$, $J_{ij}$, and
$\Delta_i$ are real-valued parameters.

One of the mechanisms that can result in a very small $g_m$ in AQO is
when an eigenstate corresponding to some local minima of $H_P$ (anti-)
crosses that of the global minimum near the end of the evolution
\cite{Amin09}. In this case, $g_m$ decreases exponentially with the
Hamming distance between the two minima. Perturbation expansion is
proven to be a useful tool to examine these problems
\cite{Amin09,Altshuler,Farhi09,DicksonPRL11}, as these anticrossings
typically happen when $\lambda = A(s)/B(s) \ll 1$. To low orders of
perturbation in $\lambda$, the perturbed energy levels cross close to
the point where an anticrossing exits in the exact spectrum of the
system. For this reason, we call this type of anticrossings, {\em
perturbative crossings}. They are also known as first order quantum
phase transition points because of the non-analyticity of the ground
state at these points in the thermodynamic limit. The position of the
antisrossings can be approximately predicted using low order
perturbation expansion, while estimating $g_m$ requires higher orders
\cite{Amin09}.

Considering random exact cover instances, it was shown \cite{Altshuler}
that the probability of having such perturbative crossings increases
with problem size, which was later numerically confirmed
\cite{Young10}. Others also found evidence of first order quantum phase
transitions, even where the perturbation expansion is expected to break
down \cite{Jorg10,Foini10,Neuhaus11}. These findings implied that
NP-complete problems may not be solved efficiently using AQO. In most
investigations on the scaling of AQO, however, at least some of the
following assumptions are quite commonly made:\\

1. $H_P$ has no free input parameters.

2. $\Delta_i$ is uniform among all qubits.

3. All minima are non-degenerate states.

4. Any suboptimal solution is an undesirable output.

5. The computation is run exactly once.

6. The system is completely isolated.\\

Assumption 1 is not always the case, as several different $H_P$ can
represent the same problem. For $H_P$ representing the Maximum
Independent Set (MIS) problem, it was shown that a significant increase
in gap size can be obtained by changing free parameters
\cite{Choi10,DicksonPRL11}. Assumption 2 was examined in
Ref.~\onlinecite{Farhi09} and it was found that choosing $\Delta_i$
values randomly can have a non-zero probability of eliminating the
anticrossing between global and local minima. Assumption 3 was
questioned by Ref.~\onlinecite{Knysh10}, who also criticized
Ref.~\onlinecite{Altshuler} based on neglecting the correlations
between the local minima.  Assumption 4 is not valid, since it is known
that some optimization problems are NP-hard to solve approximately
\cite{ApproxNPHard}. It is also commonly assumed that, at least for a closed
system, there is no benefit of running the algorithm multiple times.
For the case of open quantum systems, it has been suggested that
running the evolution faster but multiple times may lead to much higher
probability of success than running the evolution once but very slowly
\cite{Sarandy,Amin08}.

In a recent publication \cite{DicksonPRL11}, we examined the first 5
assumptions for solving MIS problems via AQO. We showed that by
changing the free parameters of the Hamiltonian, there always exists an
adiabatic path along which no perturbative crossings would occur.
However, no constructive way to find one of those paths was suggested.
In this paper, we design a heuristic algorithm to determine such a path
for a general problem. The objective is not to create a flawless
algorithm, but to demonstrate that a very simple heuristic algorithm,
similar to adaptive simulated annealing \cite{ASA}, can successfully
eliminate severe anticrossings in hard instances.

\section{Maximum Independent Set problem}

Consider a general graph ${\mathcal{G}} = (\mathcal{V},\mathcal{E})$,
with a set of nodes (vertices) $\mathcal{V}$ and a set of edges
$\mathcal{E}$. An independent set in ${\mathcal{G}}$ is a set $M
\subseteq \mathcal{V}$ such that no two of the nodes in $M$ are
adjacent. Every subset of an independent set is also an independent
set. Naturally, there are independent sets that are not subsets of a
larger independent set. We call them {\em maximal} independent sets.
Thus, an independent set is maximal if all nodes outside the set are
adjacent to at least one member of the set. The largest possible
maximal independent set, which is also the largest possible independent
set, is a {\em maximum} independent set. The problem of finding an MIS
in a general graph is called the maximum independent set problem, and
it is known to be NP-hard \cite{NPHard}.

Suppose we associate a variable $x_i \in \{0,1\}$ to each node. For a
state ${\bf x} = [x_1,x_2,...,x_n]$, we define a cost function:
 \be
 E({\bf x}) =  -\sum_{i \in \mathcal{V}} x_i
 + c \sum_{i,j \in \mathcal{E}} x_ix_j . \label{fx}
 \ee
with $c>1$. Consider an independent set $M$ of size $m$. Define a state
${\bf x}$ such that $x_i=1$ if $i \in M$, and $x_i=0$ otherwise. Since
the nodes with $x_i=1$ are not adjacent to each other, they don't
contribute to the coupling $c$-term. In other words, since for every
adjacent pair of nodes, at least one of the two has $x_i=0$, the contribution of
the coupling term to the cost function is 0. Therefore, the cost
function will be $E({\bf x}) \equiv E_M = -m$. If $M$ is a maximal
independent set, it is easy to see that $E_M$ is a local minimum of
$E({\bf x})$. This is because, by removing a node from $M$, we will
decrease $m$ and therefore increase $E({\bf x})$, and by adding another
node to it, we will increase $E({\bf x})$ by at least $c-1>0$. The
latter is because the added node will be adjacent to at least one of
the nodes inside $M$, due to the definition of maximal independent set.
Therefore, maximal independent sets are the local minima of $E({\bf
x})$ and maximum independent sets are the global minima of $E({\bf
x})$.

One can represent $E({\bf x})$ in terms of an $n$-qubit Hamiltonian by
substituting $x_i \rightarrow (1+\sigma^z_i)/2$. Ignoring a constant
energy shift, we get $H_P$ as in (\ref{HPB}) with:
 \be
 h_i = -{n_i c + 2 \over 4}, \qquad
 J_{ij} = \left\{  \begin{tabular}{cc}
 $c/4$ & \ if $i,j \in {\cal E}$ \\
 $0$ & \ otherwise \end{tabular}\right.
 \ee
where $n_i$ is the number of edges connected to (or degree of) the node
$i$.

\section{Perturbative crossings}

For perturbation expansion, it is easier to work with the re-scaled
Hamiltonian $\widetilde{H} = { H / B(s)} =  H_P + \lambda H_B$, so that
$\lambda=\infty$ for $s=0$ and $\lambda=0$ for $s=1$. The
eigenfunctions of $\widetilde{H}$ and $H$ are the same, but their
eigenvalues differ by a factor of $B(s)$. At the end of the evolution
($\lambda=0$), the eigenstates of $\widetilde{H}$ are the same as those
of $H_P$. The ground state is therefore the global minimum of $H_P$ and
the low lying excited states are either local minima of $H_P$ or states
in the neighborhood of global or local minima. At small $\lambda$, one
can use perturbation expansion to calculate the eigenvalues and
eigenstates of $\widetilde{H}$. Perturbation expansion is valid as long
as $\lambda < \lambda_c$, where $\lambda_c$ is the convergence radius
of the expansion.

Suppose there are $K$ maximal (or maximum) independent sets $M_k$ of
size $m$, with $k=1, ..., K$. States $|M_k\rangle$ and also every
superposition of them are therefore degenerate eigenstates of $H_P$,
with energy $E_M^{(0)}$. Perturbation in $\lambda$ removes this
degeneracy. Let $|M\rangle {=} \sum_k C_k |M_k\rangle$ represent the
lowest energy superposition immediately after the degeneracy is lifted.
With the positive sign of $\Delta_i$, all $C_k$ will be positive real
numbers with the constraint: $\sum_{k}C_k^2 = 1$.  Coefficients $C_k$
can be obtained by partial diagonalization of $\widetilde{H}$ in the
subspace of the local minima. The perturbed eigenvalue of this state
can be written as: $E_M(\lambda) = E_M^{(0)} + \lambda E_M^{(1)} +
\lambda^2 E_M^{(2)} + ...$, where $E_M^{(1)} = \langle M| H_B |
M\rangle$, and
 \ba
 E_M^{(2)} = \sum_{l \notin \{M_k\}} {\langle M| H_B | l \rangle \langle l| H_B |M\rangle \over
 E_M^{(0)} - E_l^{(0)}}.
 \ea
The perturbation Hamiltonian $H_B$ causes single qubit flips. The first
order correction $E_M^{(1)}=0$, because all $M_k$ are the same size
($m$), hence one cannot get from one minimum to another by a single bit
flip (adding or removing a single node). The second order correction is
 \ba
 E_M^{(2)} &=& -\sum'_{(k,k'),(i,j)}
 {\Delta_i\Delta_j C_kC_{k'} \over  B_{k,i}},
 \label{E2deg}
 \ea
where $B_{k,i}$ is the cost of flipping qubit $i$ from state
$|M_k\rangle$, and the prime sign on the sum means that the sum is over
all paths from $|M_k\rangle$ to $|M_{k'}\rangle$ with two bit flips by
first flipping qubit $i$ and then qubit $j$. This also includes $k=k'$,
which means flipping qubit $i$ two times. Notice that for positive,
real $C_k$ and $C_{k'}$, $E_M^{(2)}$ is always negative for minima.
Therefore, the second order perturbation correction always reduces the
energy of eigenstates representing minima.

Now suppose that $M$ is the unique MIS of graph ${\cal G}$ and
$|M\rangle$ represents the corresponding ground state of $H_P$ with
eigenvalue $E_M^{(0)}$. Also suppose there exist $K$ maximal
independent sets $M'_k$, $k=1,...,K$, with the same size $m'$,
producing degenerate local minima $|M'_k\rangle$ of $H_P$ with
eigenvalue $E_{M'}^{(0)}$. Equation (\ref{E2deg}), therefore, gives the
perturbed energy of the above states ($M \to M'$ for the local minima).
The two states cross at $\lambda = \lambda^*$, where the perturbed
energies are equal: $E_M(\lambda) = E_{M'}(\lambda)$. Up to the second
order perturbation, we have
$ E_M^{(0)} + \lambda^{*2} E_M^{(2)}=E_{M'}^{(0)} + \lambda^{*2}
E_{M'}^{(2)}$,
which leads to
 \ba
 \lambda^* =
 \sqrt{-(E_{M'}^{(0)}-E_{M}^{(0)})/(E_{M'}^{(2)}-E_{M}^{(2)})}.
 \ea
Since $E_{M'}^{(0)}>E_{M}^{(0)}$, in order for $\lambda^*$ to have a
real value we need $E_{M'}^{(2)}<E_{M}^{(2)}$ (or $|E_{M'}^{(2)}| >
|E_{M}^{(2)}|$ since both curvatures are negative). This means that the
local minima should have more negative curvature than the global
minimum. The magnitude of the curvature in (\ref{E2deg}) becomes large
if the energy cost $B_{k,i}$ of bit flips from the local minima
$|M_k\rangle$ is small. Moreover, if there are many degenerate local
minima with two bit flip Hamming distance from each other, each
pair of those adds 4 terms to (\ref{E2deg}). This means that if
there is a large number of local minima connected to each other by
2-bit-flip paths, they may cause a large negative curvature creating an
anticrossing with the global minimum state. We call such set of nearby
(in Hamming distance) local minima, a {\em  cluster}. In practice,
there could be many clusters of local minima and therefore there could
be many anticrossings in the adiabatic path.

Since the curvatures depend on $\Delta_i$ according to (\ref{E2deg}),
it could be possible to choose $\Delta_i$ in such a way that the
inequality $E_{M'}^{(0)}>E_{M}^{(0)}$ would not be satisfied. Thus, the
two states would not cross, at least up to second order perturbation.
Our goal here is to construct a simple iterative algorithm that
finds such $\Delta_i$ values heuristically.

\section{An iterative quantum algorithm}

Although it is difficult to determine nontrivial properties of the global minimum,
it is easy to gather information about the local minima. When there is a
perturbative crossing, if the quantum computation is run quickly
(non-adiabatically), with $t_f \ll t_a$, the system will go to the
excited state ($|M'\rangle {=} \sum_k C_k |M'_k\rangle$), after the
anticrossing, instead of the the ground state ($|M\rangle$). Repeating
this process would sample the local minima $|M'_k\rangle$ with
probability $C_k^2$.

We would like to use the above information to penalize the path to the
corresponding local minima using Eq.~(\ref{E2deg}). More specifically,
we would like to reduce the curvature $|E_{M'}^{(2)}|$ of state
$|M'\rangle$, as much as possible, by changing $\Delta_i$, without
significantly reducing that for the global minimum ($|E_{M}^{(2)}|$).
We cannot reduce all $\Delta_i$ together, because it would simply
reduce both. The bit flip energy costs $B_{k,i}$ can be easily
calculated once the local minima $|M'_{k'}\rangle$ are known via the
above sampling procedure. Determining all $C_k$, however, would require
sampling all local minima, of which there could be exponentially many.
Instead, we replace $C_{k'}$ with $C_k$ in (\ref{E2deg}), with minimal
impact on the sum on average. We obtain
 \ba
 E_{M'}^{(2)} &\approx& -\sum_i \Delta_i \sum'_{(k,k'),j} {C_k^2 \over  B_{k,i}} \Delta_j
 = -\sum_i \Delta_i \mu_i, \label{factor} \\
 \mu_i &=& \sum_k C_k^2 \left( B_{k,i}^{-1} \sum'_{k',j}
 {\Delta_j}\right). \nonumber
 \ea
Because $C_k^2$ is the probability of obtaining state $|M'_k\rangle$,
we can compute $\mu_i$ by sampling results from running quickly:
 \ba
 \mu_i = {\Bigg \langle  B_{k,i}^{-1} \sum'_{k',j} {\Delta_j}
 \Bigg \rangle}_{\text{sampled } k}. \label{mui}
 \ea
From (\ref{factor}), it is evident that $\Delta_i$ with larger $\mu_i$
contribute more to the sum, and therefore should be reduced the most.
What makes $\mu_i$ large is whether many of the degenerate local minima
can be connected by first flipping qubit $i$ and then any other qubit
$j$. When $\mu_i$ is small, $\Delta_i$ could be increased without
significantly adding to the curvature, unless its effect on other
$\mu_j$ sums in (\ref{mui}) is large.


For a given set of $\mu_i$, minimizing $\sum_i\Delta_i\mu_i$, while
keeping the geometric average $(\prod_i\Delta_i)^{1/N}$ constant,
yields $\Delta_i \propto \mu_i^{-1}$. When there is only one eigenstate
$|M'\rangle$ crossing the global minimum $|M\rangle$, this choice of
$\Delta_i$ might remove the anticrossing even after the first
iteration. However, if there exists another state $|M''\rangle$,
comprising another cluster of local minima $|M_k''\rangle$, the above
procedure may increase the curvature of $E_{M''}$, creating a new
anticrossing. This will take the system to a new set of local minima
($|M_k''\rangle$) instead of the global minimum. The second iteration
will penalize the path to $|M''\rangle$, without considering
$|M'\rangle$. This may now increase $E_{M'}^{(2)}$, which was reduced
in the first iteration, and hence we will be back to the original set
of local minima. Iterating such a process will only switch between
those two clusters of local minima. In practice, there could be more
than two clusters of local minima, so any successful algorithm must
maintain a memory of the states reached in the previous iterations.

This can be achieved by implicitly remembering previous $\Delta_i$
values, in choosing $\Delta_{i,\text{new}} \propto
\Delta_{i,\text{old}}^{1-\beta}\,\mu_i^{-\beta}$, i.e. a weighted
geometric average with the previous value of $\Delta_i$. A smaller
value of $\beta$ will adjust less and remember more, so it may take
more iterations to escape a particular cluster of local minima. A large
value of $\beta$, on the other hand, makes the system prone to getting
stuck in back-and-forth cycles between clusters of local minima. We
found that $\beta=1/(\kappa+1)$, where $\kappa$ is the current
iteration number, will converge with high likelihood. Such a $\beta$
corresponds with performing a geometric average over all previous
values of $\Delta_i$. Note that precise approximation of $\mu_i$ is not
necessary to penalize the path to a cluster of minima, so a moderate
number of samples are likely to be sufficient, and if not, the next
iteration will build on this sample.

The main algorithm is summarized in Table \ref{table}. In step 5 of the
algorithm, if some of $\Delta_i$'s are too small or too large, one can
rescale them and limit $\Delta_i$'s that are still out of range to the
acceptable minimum or maximum values.  Note that step 4 can be
completed in $O(rn^2)$ time on a single classical processor, or
$\Theta(\log(rn))$ time with $O(rn^2)$ classical processors, where $r$
is the number of samples.

 \begin{table} \caption{\label{table} The algorithm}
 \begin{tabular}{||l|l||}
    \hline\hline
1.   & \ Initialize $\Delta_i=1 \quad \forall \, i$. \\
2.   & \ Anneal $r$ times, saving each result. \\
3.   & \ If a sufficient result has been returned, {\em finish}. \\
4.   & \ Compute $\mu_i$ using (\ref{mui}). \\
5.   & \ Set $\Delta_{i,\text{new}} = \Delta_{i,\text{old}}^{1-\beta}\,\mu_i^{-\beta}$, where $\beta = 1/(\kappa + 1)$ \\
& \ and $\kappa$ is the interation number. \\
6.   & \ Rescale all $\Delta_i$'s to be within the feasible
range. \\
7.   & \ Go back to 2. \\
    \hline\hline
 \end{tabular}
 \end{table}

\section{Test Problem Instances}

In order to adequately test this algorithm in simulation, we first
required reasonably sized test problem instances with extremely small
$g_m$. Here, we focus on graphs with unique MIS (non-degenerate final
ground state), as they represent harder instances than those with
multiple MISs. Instances with a unique MIS are fairly uncommon in
uniform random graphs. In testing 80 graphs with 128 nodes and 1,572
uniform randomly placed edges, only 7 had a unique MIS, none of which
had even remotely small minimum gaps, so generating uniform random
graphs and hoping for instances with extremely small gaps would not
have sufficed for generating test instances.  Generating random graphs
from a distribution heavily skewed to have consistently small $g_m$
required a targeted approach.

A key observation is that given a maximal independent set, randomly
adding an edge between two of the nodes in the set makes the set
dependent, and creates a degenerate pair of maximal independent sets of
1 fewer node, separated by exactly 2 bit-flips.  Doing this several
times can produce many such pairs, which, as described in \cite{DicksonPRL11},
results in local minima eigenstates with large curvature as desired.

Thus, the algorithm used for generating random graphs with extremely
small gap anticrossings is:
\begin{enumerate}

\item Create a graph with 64 nodes and 220 uniform randomly
selected edges.

\item Find, by a depth-first search through the space of independent sets, an
independent set of size 20 to become the MIS, $M$.  (The expected
number of independent sets of size 20 is approximately 5.7 million, so
it is extremely likely that there is at least one.)

\item For each node, $i\notin M$ that is not adjacent to a node in $M$,
uniform randomly select a node $j\in M$, and add an edge between $i$
and $j$. This guarantees that $M$ is a maximal independent set (not MIS
yet).

\item Continue the depth-first search until another independent set of
size 20 is found. Remove one of the nodes and call this set $M'$.

\item For each node $i\notin M'$ that is not adjacent to a node in
$M'$, if $i\notin M$, uniform randomly select a node $j\in M'$, and add
an edge between $i$ and $j$; if $i\in M$, instead uniform randomly
select $j\in M' \bigcap \overline{M}$ to ensure that $M$ remains an
independent set.

\item Repeat 4 and 5 until no more independent sets of size 20 are
found.

\end{enumerate}

The last step assures that $M$ is an MIS. Caution should be used when
applying this algorithm to much larger graphs, since although it can be
executed in less than a second for 64 nodes, it does require time
exponential in the number of nodes, assuming that the desired MIS size
increases linearly with the number of nodes.

All graphs generated with the above method are guaranteed to have a
unique MIS of size 20 (unless no independent set of size 20 was found
in step 2).  Of 51 graphs generated with this method, only 1 did not
have an anticrossing with $g_m$ small enough for the test scenario below
(though it was still significantly smaller than that of all 80 of the
128-node uniform random graphs examined).  It was excluded from the
test problem set, since it would be solved immediately.

All of these generated 64-node graphs have 100,000's of maximal
independent sets (local minima), 1,000's of size 19 and 10,000's of
size 18.  Moreover, these tend to cluster into large groups of 1,000's
of maximal independent sets connected by 2 bit-flip paths. Therefore,
they are significantly harder than typical 64 qubit problems for AQO.

\section{Simulations}

We would like to simulate this algorithm for the generated problem
instances. However, because for fixed-sized systems, ``elimination" of a small gap anticrossing is not well-defined, in order to have a reasonable and objective criterion for success, we must define time and energy scales.

\begin{figure}[t]
\includegraphics[trim=2.5cm 2.6cm 2.3cm 2.4cm,clip,width=7cm]{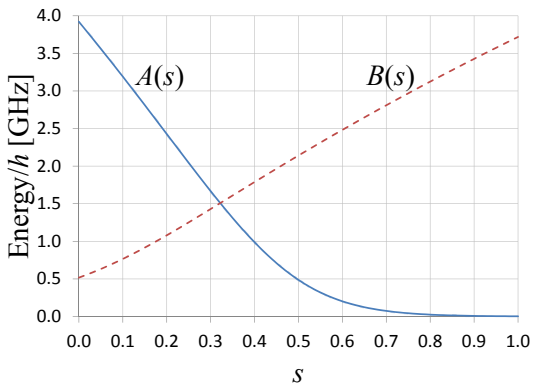}
\caption{(Color online) $A(s)$, the energy scale of $H_B$, and $B(s)$,
the energy scale of $H_P$.} \label{AandB} \end{figure}

We choose as our $A(s)$ and $B(s)$, energy scales extracted from
superconducting flux qubits similar in design to those examined in
\cite{Harris10,hramping}, which are plotted in Fig. \ref{AandB}. We
choose $r$, the number of times the quantum computation is performed
per iteration, to be 500, and the annealing time for each such
computation to be $t_f = 0.08~\mu$s.  Thus, if $t_a <16~\mu$s, there is
a high probability ($>92$\%) that at least one of the 500 results from an
iteration will be the global minimum. Also, for any result obtained
that is not a minimum, one can easily perform gradient descent to reach
a minimum, possibly the global minimum.  Thus, if in an eigenstate
crossing the ground state, the total probability of states that descend
to the global minimum is $>0.005$, there is a high probability
($>92$\%) that the global minimum can be obtained in this manner.
Either case is considered to be successful.

\begin{figure}[t] \includegraphics[trim=3cm 1.2cm 16cm
1.2cm,clip,width=5.5cm]{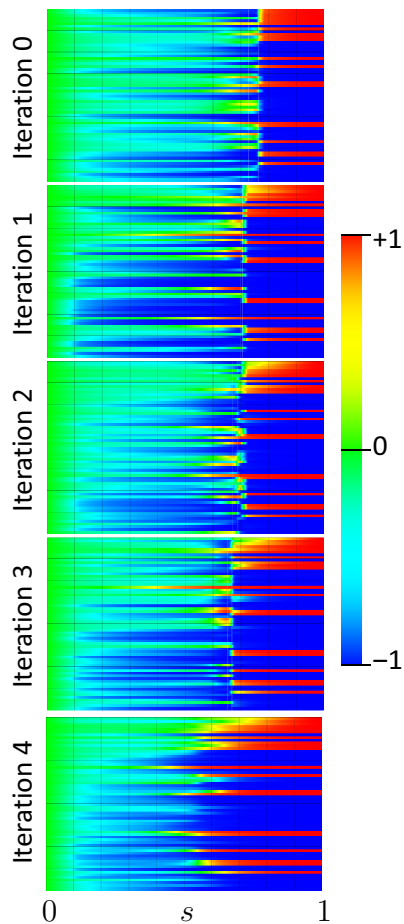} \\
 {\large 0 \hspace{1.35cm} $s$ \hspace{1.35cm} 1 \ }
\caption{(Color online) Visualization of perturbative crossings
eliminated in 4 iterations. Each horizontal strip plots the ground
state expectation $\langle \sigma_z^{(i)}\rangle$, for the $i$th qubit,
as a function of $s$. There are 64 strips in each plot corresponding to
the 64 qubits. All qubits begin in a uniform superposition, so
$\langle\sigma_z^{(i)}\rangle = 0$ (green) at $s=0$ on the left. The
final ground state represents the MIS, with $+1$ (red) for nodes in the
set, and $-1$ (blue) for nodes not in the set. As $s$ increases from 0
to 1, the system localizes into low-energy minima (green moving toward
red or blue).  If it localizes into the global minimum, like the smooth
transition in iteration 4, there is no perturbative crossing. However,
if it localizes into local minima, there is at least one crossing,
which will be visible as a sudden change in many qubits, as seen in
iterations 0-3.  Each iteration penalizes the path to the local minima
into which the ground state localized, until no crossings remain.  The
penalization is strong enough that different local minima are found on
each iteration.} \label{EliminationVisual}
\end{figure}

As described earlier, if the minimum gap is very small, such that $t_f
\ll t_a$, the system will occupy the excited state $|M'\rangle$ after
the anticrossing and one of the local minima will be reached, and the
algorithm uses a sampling of these minima to calculate $\mu_i$ using
(\ref{mui}). Here, however, we use Quantum Monte Carlo (QMC)
simulations \cite{Young10} to provide the sampling needed. QMC provides
samples of computation basis states in the proportions that they appear
in the ground state of the system. Before a small gap anticrossing, the
ground state is approximately the superposition state
$|M'\rangle=\sum_k C_k |M_k'\rangle$. Therefore, using QMC to sample
just before the anticrossing gives samples approximately as they would
come from evolving the system.  We use this fact to calculate $\mu_i$
in (\ref{mui}). By incorporating gradient descent into the sampling, we
can also determine the occupation of states in the well of the global
minimum, in addition to correcting for some of the single bit flip
deviation from $|M'\rangle$. The adiabatic time (\ref{ta}) can be
computed using QMC in the same manner as described in \cite{Karimi}.
If on an iteration, the adiabatic time is small enough to be
computed accurately and is found to be $<16~\mu s$, or the probability
of states in the well of the global minimum just before a perturbative
crossing is $>0.005$, the instance is considered solved on that iteration.  This is because the probability of finding the global minimum at least once in the 500 computations of the iteration is then $>92$\%.

The ``feasible range" of $\Delta_i$ values, as mentioned in the
algorithm description, was chosen to be between $1/4$ and $8$.  More
specifically, on each iteration, the $\Delta_i$ values would always be
scaled such that the smallest was $1/4$, and the other $\Delta_i$
values rarely approached $8$, especially after several iterations,
where most $\Delta_i$ values were $<4$.

\begin{figure}[t] \includegraphics[trim=2.2cm 3cm 2.5cm
3cm,clip,width=7cm]{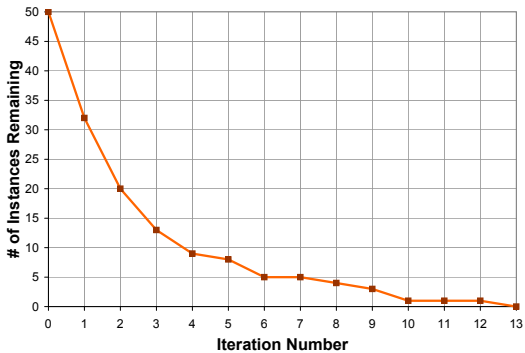} \caption{(Color online) Number
of ``unsolved" problem instances as a function of the number of
iterations. All instances were solved within 13 iterations.}
\label{RemainingInstances} \end{figure}

Beyond just testing whether the algorithm presented here is successful
or not, it is critical to gain insight into how the changes in
$\Delta_i$ values affect the evolution of the ground state.  Although a
$2^{64}$-dimensional system cannot feasibly be examined in detail, much
can be seen by examining the ground state expectation values
$\langle\sigma_z^{(i)}\rangle \equiv \langle 0|\sigma_z^{(i)}|0
\rangle$ of the 64 operators, i.e., the average magnetization of each
qubit in the instantaneous ground state. Figure~\ref{EliminationVisual}
illustrates these expectation values as a function of $s$ for 4
iterations of the algorithm with an example instance. In each of the 4
plots, 64 horizontal strips color-code the value of
$\langle\sigma_z^{(i)} \rangle$ for all 64 qubits, during the
evolution, calculated using QMC sampling. At $s=0$, the ground state is
a uniform superposition of all computation states, so
$\langle\sigma_z^{(i)} \rangle=0$ (green), and at $s=1$, the ground
state is the global minimum of $H_P$, therefore $\langle\sigma_z^{(i)}
\rangle=\pm 1$ (red/blue) depending on the state of the qubit. Moving
away from $s=0$, the expectations gradually tend toward $-1$ or $+1$ as
the ground state settles into global or local minima. If the system
directly settles into the global minimum (as in iteration 4), then
there will be a continuous change of $\langle\sigma_z^{(i)} \rangle$
from 0 to their ultimate values with no sharp transition. If, on the
other hand, the system initially settles into some cluster of local
minima, then a sudden change is expected at the anticrossing between
the eigenstates corresponding to the local and global minima
($|M'\rangle$ and $|M\rangle$). Iterations 0-3 clearly show such sudden
transitions. After each iteration, the anticrossing moves to an earlier
time, though this was not general among all the instances examined. In the last iteration, the anticrossing is completely removed and
the transition from the beginning to the end state is smooth.

%

All 50 of the extremely difficult 64-node random instances were solved
within 13 iterations of the algorithm. The number of unsolved instances
after each iteration is plotted in Fig.~\ref{RemainingInstances}. Among
the instances, 30 of them were solved in 2 iterations, and all but 1
were solved in 10 iterations, with an overall average of 3.0 iterations
required. These data suggest that the selection of $\beta$ to ensure
equal application of penalties is quite robust at remembering previous
penalties while still applying new penalties.

\section{Conclusions}
We have demonstrated that a simple adiabatic quantum algorithm, based
on penalization of paths to clusters of local minima by tuning
single-qubit tunnelling energies, is effective at eliminating extremely
small gaps caused by perturbative crossings. We presented a method for
generating 64-qubit random instances of maximum independent set with
$10^5$ to $10^6$ highly degenerate local minima and a unique global
minimum, causing perturbative crossings between the two.  It is found
that even for these instances, the algorithm can eliminate the
perturbative crossings in a small number of iterations.

\subsection*{Acknowledgements}
The authors are grateful to B.~Altshuler, P.~Bunyk, E.~Chapple,
P.~Chavez, E.~Farhi, S.~Gildert, F.~Hamze, R.~Harris, M.~Johnson,
T.~Lanting, K.~Karimi, H.~Katzgraber, T.~Mahon, T.~Neuhaus,
R.~Raussendorf, C.~Rich, G.~Rose, M.~Thom, E.~Tolkacheva, B.~Wilson,
and A.P.~Young for useful discussions.  The authors also thank the volunteers of the AQUA@home project, who donated their computing resources to run the QMC simulations for this work.

\end{document}